\def\beginenum{\begin{enumerate}}
\def\endenum{\end{enumerate}}
\def\bgary{\begin{array}}
\def\edary{\end{array}}
\def\bgeqns{\begin{eqnarray}}
\def\edeqns{\end{eqnarray}}
\def\bgeq{\begin{equation}}
\def\edeq{\end{equation}}
\def\alf{\alpha}

\def\lam{\lambda}
\def\ep{\epsilon}

\def\k{{\vec k}}
\def\q{{\vec q}}

\documentstyle [12pt]{article}
\begin{document}
\title{Nonequilibrium Effects and Self Heating in Single Electron Coulomb
Blockade Devices}
\author{Chiu Liu and Qian Niu\\
Department of Physics \\
The University of Texas at Austin\\ Austin, Texas\\  78712}
\maketitle
\begin{abstract}

We present a comprehensive investigation of nonequilibrium effects and self
heating
in single electron transfer devices based primarily on the Coulomb blockade
effect.
During an electron trapping process, a hot electron may be deposited in a
quantum dot or metal island, with an extra energy usually on the order of the
Coulomb charging energy, which is much higher than the temperature in typical
experiments.
The hot electron may relax through three channels: tunneling
back and forth to the feeding lead (or island), emitting phonons, and exciting
background
electrons.  Depending on the magnitudes of the rates in the latter two channels
relative
to the device operation frequency and to each other, the system may be in one
of
three different regimes: equilibrium, non-equilibrium, and self heating
(partial equilibrium).
In the quilibrium regime, a hot electron fully gives up its energy to phonons
within a pump cycle. In the nonequilibrium regime, the relaxation is via
tunneling with a
distribution of characteristic rates; the approach to equilibrium goes like a
power law of time
(frequency) instead of an exponential.  This channel is plagued completely in
the continuum limit of
the single electron levels.   In the self heating regime, the hot electron
thermalizes
quickly with background  electrons, whose temperature $T_e$ is elevated above
the lattice temperature
$T_l$.  We have calculated the coefficient in the well known $T^5$
law of energy dissipation rate, and compared the results to experimental values
for aluminum and copper islands and for a two dimensional semiconductor quantum
dot.
Moreover, we have obtained different
scaling relations between the electron temperature and operation frequency and
device size
for various types of devices.

\end{abstract}
\par

\par
\centerline {PACS Indices:  73.40.Gk, 06.20.Hq, 72.10.Di, 72.15.Lh, 73.20.Dx}
\tableofcontents
\endtitlepage

\section{Introduction}
\paragraph{}
Electron turnstiles and pumps have been made using metal
islands [1-4] and semiconductor quantum dots [5,6] based on
the Coulomb blockade effect. Other types of single electron transfer
devices have also been proposed [7-9]. These devices have promised
great potential in applications to high precision metrology (charge or current
standard)
and computer technology (memory and logic circuits) \cite{Lik}.
To date, several papers have been devoted to the study of possible precision
of such devices, with various sources of error identified, such as thermal,
cotunneling,
nonadiabatic, and noise effects [11-16].

\paragraph{}

In this article, we present a comprehensive investigation of non-equilibrium
effects
 and self heating in single electron transfer devices which are primarily
based on the Coulomb blockade effect.  In such devices,  the electron
temperature
is very low compared to the Coulomb charging
energy, but may exceed the level spacing in the excitation spectrum.
Devices based on metal islands clearly lie in this regime, and this is also
true
for currently available quantum dot turnstiles or pumps, where excitation level
spacing
is typically several times smaller than the Coulomb charging energy.  For ease
of
reference, we will use the term {\it Coulomb island} for a metal island or
quantum dot
in such a regime.
Some highlights of this investigation have previously been published
\cite{Liu.P,Niu.P}.
The primary purpose of this article is to provide a complete picture of the
different regimes,
with comprehensive arguments and calculations included.  A number of new
results
will also be presented.

\paragraph{}

We first examine a quantum dot electron turnstile to identify a major source of
non-equilibrium and self heating: the deposition of hot electrons in a Coulomb
island
\cite{hot}.
This process is common to all Coulomb blockade devices.
Figure 1 shows the process of an electron tunneling from a lead into the
quantum dot,
which occurs when the barrier between them is lowered \cite{Kou:PRL,Kou:Z}.
The bias is large enough to overcome the coulomb charging energy ($U$) for the
addition
of one electron into the dot, but not so large as to allow a second electron to
enter the dot.  To minimize the thermal probability of adding no electron or
more than one electron into the dot, the Fermi energy of the feeding lead is
typically placed at about $U/2$ higher than the lowest available state in the
dot
before tunneling.  There are then  several available states into
which an electron can tunnel from the lead.  It is imaginable that the electron
may actually tunnel into a state of about $U/2$ higher in energy than the
lowest available
state in the dot, the probability of which is further enhanced by the fact that
the
tunneling rate increases rapidly with energy.
\paragraph{}

There is a similar non-equilibrium process during the draining part of a cycle.
The Fermi energy of the draining lead is usually biased many levels below the
highest occupied one.  When an electron tunnels out to the draining lead, it
may come
from a level below the highest occupied one.   As a result, a hot hole may be
stuck in the
dot or the metal island for a long time due to the weak coupling to the lead
and to the
phonons and background electrons at low temperatures.
Since the hot hole effect is entirely similar to the hot electron effect, we
will not
consider it explicitly for  most   of this article.

\paragraph{}
We then consider an electron turnstile based on small metal islands as shown in
Figure 2.
Unlike the quantum dot device, the tunneling barriers cannot be varied, and
electron transfer
is purely controlled by changing the various potential biases in the circuit.
To trap an electron in the middle island from the left lead,
a bias larger than $e/C$
is applied between them, where C is the capacitance of the tunnel junctions.
One possibility is that an electron will enter from the lead into the left
island,
causing the Fermi energy of the left island to be $U\equiv e^2/2C$ higher than
the
lowest available state in the middle island.  As a result, a hot electron will
be deposited
into the middle island from the left island.  An alternative possibility is
that an electron
tunnels first from the left island to the middle island followed by the
deposition of a hot electron
in the left island from the left lead.  Finally, there is the possibility of
co-tunneling,
which may also result in a hot electron in the middle island \cite{Ave-co}.

\paragraph{}
The situation in an electron pump is slightly different (Figure 3).  The
tunneling through a junction
occurs while the bias across it is ramped up through the threshold of Coulomb
blockade.
In the adiabatic limit,
tunneling will occur at the threshold, with one electron deposited in a lowest
available level in the final island.  However,
a hot electron will be in the final states, if the ramping rate is finite.  On
average,
the extra energy of the hot electron increases as a square root of the ramping
rate \cite{Jensen}.

\paragraph{}
Therefore, a non-equilibrium situation can easily arise in a Coulomb island,
unless the hot electron can relax to the lower available states in a time short
compared to
the operation cycle of the device.  There are three ways of hot electron
relaxation:
(1) tunneling back and forth, (2) emitting phonons, and (3) thermalizing with
other electrons.
In the first process, electrons tunnel back and forth between the Coulomb
island
and the feeding lead (or island), eventually bringing the two regions into
mutual equilibrium.
However, such a relaxation process can be extremely slow at low temperatures,
because the
rate of tunneling in a second electron is severely reduced by the Coulomb
blockade, and that
of tunneling out the hot electron is very effectively blocked by the
lack of holes in the Fermi sea of the feeding lead (island).
In the second process, the hot electron simply emits a phonon and makes
a transition to a lower state.  At typical energies of order 0.3 meV or lower,
only acoustic phonons are emitted, and the rate of emission is significantly
limited by the
weakness of electron-phonon coupling and the lack of phase space.
Finally, in the third process,  the hot electron relaxes at the expense of
exciting other
electrons already present in the island.  The hot electron may reach
equilibrium with others
in the island by electron-electron scattering, but the island as a whole is
heated.

\paragraph{}
There are three regimes of operation according to how fast the rates of the
relaxation processes are
compared to one another and to the operation frequency of a device.
These are illustrated in Figure 4
as different regions in the $\gamma_{ee}$-$\gamma_{eph}$ parameter space, where
$\gamma_{ee}$ is the
characteristic rate of the hot electron relaxation due to interaction with
other electrons,
and $\gamma_{eph}$ is the rate due to emitting phonons.  When both rates are
slow compared to
the operation frequency $\nu$ of the device, we have the non-equilibrium
situation.
When the phonon emitting rate $\gamma_{eph}$ is larger than both the operation
frequency
$\nu$ of the device and the thermalizing rate $\gamma_{ee}$, we have the
equilibrium situation.
On the other hand, when the thermalizing rate $\gamma_{ee}$ is larger than both
$\nu$ and
$\gamma_{eph}$, we have the self heating (partial equilibrium) regime.

\paragraph{}
The article is organized according to the different regimes.
The non-equilibrium regime is studied in detail
in Section 2, where only the tunneling process is taken into account.  One will
see
that the relaxation of a hot electron is characterized by a set of distributed
rates,
the lower spectrum of which are severely frozen at low temperatures.   As a new
result, we
will establish a power law frequency dependence of energy relaxation of the hot
electron when the
level spacing  in the Coulomb island becomes smaller than the temperature.
Some
details of this section are included in the Appendix.

\paragraph{}

In order to delimit the equilibrium regime, we calculate in Section 3 phonon
induced interlevel
transition rates, and study how the relaxation rates found in the previous
section are affected by
the phonons under experimental conditions.  Discrete level effects are
emphasized for applications to
quantum dots, and the continuum  Fermi liquid model is considered for metal
islands.

\paragraph{}

Section 4 is devoted to the study of the self heating regime.
In the first part of this section, $\gamma_{ee}$ is estimated for typical
quantum dots and metal
islands.
In the second and third parts, the rates of energy dissipation from thermalized
electrons
to phonons are obtained for a metal island and a quantum dot, respectively.
The coefficient in the $T^5$ law of energy dissipation is calculated for
aluminum and copper
islands and for a two dimensional quantum dot, and these new results are
compared to experimental
values. In the last part of Section 4, we show how the electron temperature and
thermal error
scale with the frequency and size for different devices.  Finally, we summarize
in Section 5.

\section{Hot Electron Relaxation Via Tunneling}
\paragraph{}
In this section, we simulate numerically the action of an electron turnstile,
and show
analytically how a hot electron can appear and relax in a Coulomb island during
the
trapping part of an operating cycle.  As mentioned in the introduction, we will
concentrate our attention on the tunneling processes,  with studies of the
effects of phonons and
background electron excitations deferred to the next two sections.  Our
discussions will be
mostly focused on a quantum dot device with discrete levels, but we will also
consider the
case where the level spacing is small compared with temperature in the leads,
and compare
the results with existing theory about tunneling between metal islands.

\subsection {Modeling the Quantum Dot Electron Turnstile}
\paragraph{}
Our model consists of a quantum dot separated from two leads by tunable energy
barriers.
The leads are represented by
two electron seas with Fermi energies $E_f^L$ and $E_f^R$ respectively.
We will consider a set of $N$ available single electron levels with spacing
$\Delta$
in the dot.  We place $E_f^L$
in the middle of the gap above the $[{N\over2}]$th level (counted from below),
and  $E_f^R$ at $U/2$ below the lowest available level, where $U$ is the
Coulomb charging
energy in the dot \cite{Meir,Been}.
Corresponding to current experimental conditions, we will assume
 $\hbar \nu\ll k_BT \ll U$ and $\hbar\Gamma<<\Delta$,
where $\nu$ is the operating frequency and
$\hbar\Gamma$ the level width due to tunneling.
Then semiclassical rate equations can be used
to describe the tunneling processes:
\bgeq
{dP_{\alpha} \over dt}= \sum_{l=L,R} g_{\alpha}^l(t)
[(1-{\sum_{\alpha'} P_{\alpha'}})f_{\alpha}^l-P_{\alpha}(1-f_{\alpha}^l)],
\edeq
where $f_{\alpha}^l=[1+\exp({E_\alpha-E_f^l \over k_BT})]^{-1}$ is the
Fermi-Dirac distribution
function in the leads,
$P_{\alpha}$ is the level occupation probability in the quantum dot,  and
$g_{\alpha}^l$ is the
coupling of the dot to the leads.  The factor $(1-{\sum_{\alpha'}
P_{\alpha'}})$ in the
tunneling-in term reflects the Coulomb blockade effect in the large ${U\over
2k_BT}$
limit.  The rate equations have been applied to many different situations,
such as the study of conductance oscillations
in a quantum dot device \cite{Van,Been}, single electron tunneling through
metal junctions
\cite{Aver},
charge transfer scattering of atoms or ions from a metal
surface \cite{Blan,KO} and non-equilibrium effects in a turnstile dot
device \cite{Liu.P}.  In the Appendix, we give a brief summary of a study on
the
non-Markovian correction to the above rate equations
\cite{Liu.dis,Bruder}, showing that the semiclassical rate equations are
accurate enough even
for the purpose of examining  exceedingly small errors of electron trapping and
transfer.

\paragraph{}
The couplings between the states in the quantum dot and the leads are modeled
as follows.
To simulate the sensitive energy dependence of the couplings, we take
$g_\alpha^l(t)=g_f^l(t) a^{\alpha-\alpha_f}$, where $a>1$ is a constant, and
$\alpha_f$
is the label of the state immediately below $E_f^L$ \cite{Meir}.
To move one electron from lead $L$ to lead $R$ per cycle,
we adopt the time-periodic couplings as
$g^L_f(t)=\Gamma^L_f e^{-\delta(1+\cos(\omega t))}$
and  $g^R_f(t)=\Gamma^R_f e^{-\delta(1-\cos(\omega t))}$,
where  $\delta$ characterizes the variation
amplitude of the coupling strengths. It can be seen that in the first half of
the
cycle, the coupling to the right is negligible and only those electrons in the
left
lead is allowed to move into the dot; while in the second half, the left lead
is nearly
blocked and the trapped electron moves to the right via the large coupling to
the
right lead.  For simplicity, we assume that the energy levels do not change
with time.

\subsection{Trapping Probability and Error}
\paragraph{}
The basic operation principles of all single electron transfer devices are
based on
an adiabatic analysis appropriate for very low operation frequencies.
In the adiabatic limit,
the time dependence of the
occupation probability of the dot, $P(t)$, is found by letting the right hand
side of Eq.(1)
vanish, yielding  $P(t)={S}{[1+ S]}^{-1}$ with
$S=\sum_\alpha f_\alpha^L
[ (1-f_\alpha^L)+{g_f^R(t) / g_f^L(t)}
]^{-1}$,
where we have dropped the terms of order $f_\alpha^R$ under the assumption
that $e^{-{U\over 2k_BT}}$ is as small as desired.
At $t=0$, where $g_f^R/g_f^L$ takes the maximum
value of order $e^{2\delta}$, $S$ and therefore $P(t)$ take a minimum value
of order $e^{-2\delta}$. This is due to leakage through the left barrier which
is
high but finite at t=0.
Since we are only interested in non-equilibrium effects due to finite
operation frequencies, we assume that the leakage error is also as
small as desired.
On the other hand, at $t=\pi/\omega$, where $g_f^R/g_f^L$ takes the minimum
value of $e^{-2\delta}<<1$, $S$ is of order of
$(e^{-U/2k_BT}+e^{-2\delta})^{-1}$, and $P(t)$ approaches
unity with a deficiency of the order of $(e^{-U/2k_BT}+e^{-2\delta})$, which is
to be
ignored according to the above discussion.

\paragraph{}
We have numerically solved for the occupation probability $P(t)$ and
tunneling currents as functions of time for finite operating frequencies.
Shown in Figure 5 is the case for $\nu=20$ MHz, with realistic parameters taken
from
Ref.\cite{Kou:PRL,Kou:Z}
as $\delta=10$,
${\Delta \over k_BT}=4$, $a=1.5$, $\Gamma_f^L=\Gamma_f^R=10^{10}$ /s and
$N=10$.
The time interval shown is taken after the first cycle so that the memory
effect of
the initial condition is absent.
The general trend in $P(t)$ follows the adiabatic result, but the
minimum point $t_0$  and the maximum point $t_m$ are shifted by $0.2\pi$ above
$\omega t=0$ and by $0.32\pi$ above $\omega t=\pi$,  respectively.
The tunneling currents from the
left lead and to the right lead are given by $I_l=\pm g_{\alpha}^l(t)
[(1-{\sum_{\alpha'} P_{\alpha'}})f_{\alpha}^l-P_{\alpha}(1-f_{\alpha}^l)]$,
where
$l=L, R$. They are shown in Figure 5 to aid our understanding of the
electron trapping and expelling processes.
It is noticed that when $P(t)$ rises or drops down to $1\over2$ at
$t_{1/2}={0.72\pi}$ or ${1.72\pi}$, the current $I_L$ or $I_R$ reaches its
maximum.
In the interval of $(t_0,t_m)$  the unwanted leakage $I_R$ is negligibly small
($<8\times 10^{-6}\omega$),
this corresponds to the trapping process;  while in the interval
$(t_m,t_0+1/\nu)$ the leakage
$I_L$ is negligibly small ($<9.8\times10^{-6}\omega$), this corresponds to the
expelling process.
The probability that an electron is transferred from the left to the right lead
in a time cycle is
$\int_{t_0}^{t_0+1/\nu} I_L\approx\int_{t_0}^{t_m} I_L\approx Max(P)-Min(P)$,
where in the first approximation we have neglected $I_L$ in the
second half of the cycle, and in the second approximation we have neglected
$I_R$ in the first half.  Also, we found that $Min(P) \sim 10^{-6}$.
Therefore, up to terms of order $10^{-5}$, the electron
transfer probability is the same as the electron trapping probability $Max(P)$
in the first half cycle.

\paragraph{}
We would like to know how the trapping error varies as a function of the
operation
frequency.  In Figure 6, the trapping error $(1-Max(P))$
(open diamonds) is plotted against $1/\nu$ on log-log scales.
Except for the very lower end, the curve
also represents the transfer error according to the above discussions.
The general decreasing trend of the curve
appears like a power law decay, but
the downward curving segments are actually closely represented by exponentials.
We explain these points in the following analysis.

\subsection{Distributed Relaxation Rates}
\paragraph{}
To explain the frequency dependence of the
trapping error, we closely examine the time
interval ${\pi\over 2}< t<t_m$ (see Figure 5), in
which the coupling to the right lead can be ignored.  We can then write
\bgeq
 {dP_{\alpha} \over dt}=g_f^L(t) a^{\alpha-\alpha_f}[
(1-{\sum_{\alpha'} P_{\alpha'}})f^L_{\alpha}-P_{\alpha}(1-f^L_{\alpha})].
\edeq
We first find
the equilibrium (adiabatic) solution $P_\alpha^{eq}$
of the equation by letting the right hand side equal zero, yielding:
$P_\alpha^{eq}= {e^{{E_{f}^L-E_{\alpha}} \over {k_BT}}}
(1+ S)^{-1}$,
where $S =\sum_\alpha e^{E_f^L-E_\alpha \over k_BT}$.
The equilibrium occupation numbers obey the Boltzmann's distribution,
and the sum $\sum_\alpha P_{\alpha}^{eq}$ approaches unity
with a negligible correction $\sim e^{-{U}\over{2k_BT}}$.

\paragraph{}
Next, we consider the actual probability distribution, and see how it differs
from
equilibrium.  For the time interval ${\pi\over 2}<t<t_{1/2}$, we may assume
that
the occupation of each level is negligible, so that the term proportional to
$P_\alpha$ on the right hand side of Eq.(2) may be dropped.  We then find that
$P_\alpha(t)$  is proportional to $a^{\alpha -\alpha_f} f_\alpha^L$,
with the $\alpha$-independent coefficient equal to
$\int_{\pi\over 2}^t dt' g_f^L(t')[1-P(t')]$,
where we have neglected the initial value $P_\alpha({\pi\over 2})$.
The coefficient can be fixed to the more explicit form,
$P(t)/\sum_\alpha a^{\alpha -\alpha_f} f_\alpha^L$,
by using the condition that $P_\alpha$ should sum up to $P$.
It is clearly seen that the actual probability distribution in this time
interval is
far from the equilibrium Boltzmann distribution.  In fact, it is peaked at the
level just below the Fermi energy, corresponding to the tendency of depositing
a hot
electron there.  This distribution is well reproduced in the numerical
calculation even
at $t=t_{1/2}$.

\paragraph{}
To see how this distribution relaxes to equilibrium in the
interval, $t_{1/2}<t<t_m$, we have performed an eigenvalue analysis.
The deviations, $\Delta P_\alpha(t)$, from the equilibrium occupation
numbers may be written as
\bgeq \Delta P_\alpha(t)=\sum_j A_{\alpha}^j e^{-\lambda_j \int_{t_{1/2}}^t
g_f^L(\tau)d\tau}, \edeq
where  the $A$'s and $\lambda$'s
are eigenvectors and eigenvalues (labeled by $j$) of the following equation
\bgeq
\lambda A_\alpha=a^{\alpha-\alpha_f}[
({\sum_{\alpha'} A_{\alpha'}})f^L_{\alpha}-A_{\alpha}(1-f^L_{\alpha})].
\edeq
The eigenvectors are only determined up to normalization factors, but they can
be
fixed by using the values of $\Delta P_\alpha(t_{1/2})$, which are known from
the
discussions in the preceding two paragraphs.
Up to terms of order $e^{-{U}\over{2k_BT}} $, the trapping error $1-Max(P)$ is
just the negative of
$\sum_\alpha \Delta P_\alpha(t_m)$, and is given by
$\sum_j B^j e^{-\nu_j/\nu}$,
where $B^j=-\sum_\alpha A_\alpha^j$ and $\nu_j=\lambda_j q(t_m)\nu \approx
\lambda_j\Gamma_f^L/\sqrt{2\pi\delta}$, which is frequency-independent.
The trapping error calculated this way is
plotted in Figure 6 as the solid curve,
which fits the result of two-lead numerical simulation, represented by the
open diamonds, very closely for
the entire range of several decades of data.

\paragraph{}
The eigenvalues are listed in Table 1 for the parameters $N=10$,
$a=1.5$, and  $\Delta/k_BT=4$, which are used for Figure 6.
Note the huge differences in the magnitude scales of the eigenvalues.
At $\delta=10$ and $\Gamma_f^L=10^{10}$ /s as used in the figure,
they give rise to the following series of characteristic relaxation rates:
$\nu_1=0.13\times 10^3$ /s, $\nu_2=14\times 10^3$ /s, $\nu_3=1.2\times 10^6$
/s, $\nu_4=96\times
10^6$ /s, and etc.  The three exponential segments seen in the figure are
attributed to $\nu_2$ to $\nu_4$.  One can also predict that
at very low frequencies, the exponent should be given by
$\nu_1$.  The power-law-like
overall trend of the curve is a result of distributed exponentials.

\paragraph{}
We now present an analysis of the eigenvalues in order to
see their physical origin.  If we solve for $A_\alpha$ from
the eigenvalue equation, and make a summation over $\alpha$,
we find that $\lambda$ satisfies the equation
${\sum_{j=1}^{N}}{b_j \over {\lambda -d_j}}=1$, where
$d_j=a^{j-\alpha_f}(1-f_j^{L})$ and $b_j= a^{j-\alpha_f}f_j^{L}$.
It can be seen that $\lambda_j$ lies in between $d_j$ and $d_{j+1}$
for $j=1,...N-1$, and that $\lambda_N$ lies above $d_N$.
All the eigenvalues are positive as required.  Because of the
Pauli exclusion factor $(1-f_j^{L})$, the $d_j$'s with $j<\alpha_f$
are exponentially small at low temperatures, which, together with the
fact that the tunneling rates increases rapidly with level energy ($a>1$),
explains why the lower eigenvalues become
so small. Quantitatively, each of the
roots can be approximately found by keeping only the neighboring terms in the
summation series, yielding $\lambda_j={{a^{(j-\alpha_f+1)}(1-f_{j+1})}\over
{1+a}}$ for $1\le j<\alpha_f$, and $\lambda_{j}\sim a^{j-\alpha_{f}}$ for those
$j\ge\alpha_f$. These rough estimates agree well with the numerical
calculation.

\paragraph{}
We have  also done a similar calculation for ${\Delta\over k_BT}=2$, and
found that  the characteristic frequency scales are raised to $\nu_1=0.15\times
10^6$ /s,
$\nu_2=2\times 10^6$ /s, $\nu_3=24\times 10^6$ /s, $\nu_4=227\times 10^6$ /s,
and etc..  The
qualitative behavior may  be understood by the fact that the Pauli exclusion
effect is less strong at
a higher  temperature.  Finally, we note that the characteristic frequency
scales may also be
increased (in our favor) by increasing $\Gamma_f^L$ and lowering $\delta$, but
one
must pay the  price of increasing the unwanted leakage.

\subsection {Power Law of Hot Electron Relaxation}
\paragraph{}
We mentioned in subsection 2.2 that the overall trend of the trapping error
looks like a power law of frequency.  To explain this, we consider in this
subsection
the case of small level spacings compared to the temperature.
We expect that the exponential segments in
Figure 6 will become invisible, and that the frequency dependence of the
trapping error will become a
pure power law.  Such a study is particularly relevant to devices based on
metal islands
[1-4],
where the single particle levels form a continuum.

\paragraph{}
The major tool of our analysis is the method of Laplace transformation.  We
first
rewrite Eq.(2) as
\bgeq
 {dP_{\alpha} \over ds}= a^{\alpha-\alpha_f}[
(1-{\sum_{\alpha'} P_{\alpha'}})f^L_{\alpha}-P_{\alpha}(1-f^L_{\alpha})],
\edeq
where $s=\int_{\pi/(2\omega)}^t g_f^L(t') dt'$. Performing a Laplace
transformation with respect to
this new `time' variable, we obtain that $\lambda {\tilde
P_\alpha}=a^{\alpha-\alpha_f}
[(1/\lambda-{\tilde P})f_\alpha^L -{\tilde P_\alpha} (1-f_\alpha^L)]$,
where $\lambda$ is the Laplace variable and $\sim$
denotes transformed quantities.  We have used the initial condition that the
probabilities are
zero.   Solving for $\tilde P_\alpha$ in terms of $\tilde P$ and summing over
$\alpha$ yields
an algebraic equation for $\tilde P$, which has the solution $\tilde P={1\over
\lambda}
{S\over 1+S}$, where $S=\sum_\alpha {f_\alpha^L a^{\alpha-\alpha_f}\over
\lambda+
(1- f_\alpha^L)a^{\alpha-\alpha_f}}$.  $P(s)$ is then obtained from the
standard formula
of inverse Laplace transformation: $\int_c {d\lambda\over 2\pi i}e^{\lambda s}
\tilde P(\lambda)$,
where the contour surrounds the singularities of $\tilde P(\lambda)$.

\paragraph{}
The pole at $\lambda=0$ yields the equilibrium probability $P_{eq}={S(0)\over
1+S(0)}$,
which is the same as before.
The other singularities are poles lying on the negative part of the real axis,
which become a
branch cut along the segment $(-1, -h_0)$ in the limit of densely distributed
levels,
where $h_0=1-f_0^L<<1$ corresponding to the lowest available level.   In the
asymptotic region
$1<<s<<1/h_0$, we may evaluate the trapping error analytically as
\bgeq
P_{eq}-P(s)={\Delta\over  k_BT} {\Gamma(\eta)\sin^2 {(1-\eta)\pi}
\over \eta \pi^2}s^{-\eta}, \label{er-emt}
\edeq
where $\eta=(1+{k_BT\over \Delta}\ln a)^{-1}$, $\Delta$ is the level spacing,
and $\Gamma()$ is the standard
$\Gamma$ function.
The case of $a=1$ ($\eta=1$) needs special consideration, where the second
factor on the
right hand side of the above
equation should
be replaced by the logarithm $(\ln sh_0)^{-2}$; however, in general, $a>1$ and
$\eta<1$ holds
since the coupling to a dot or a metal island decreases as the level energy
drops.

\paragraph{}
Since at $t=t_m$ we have $s=\Gamma_f^L/(\nu\sqrt{2\pi\delta})$,
the above asymptotic range corresponds
to $\nu_0h_0<<\nu<<\nu_0$, where $\nu_0=\Gamma_f^L/\sqrt{2\pi\delta}$.
The above result says that in this range
the trapping error goes as a sub-linear power law of frequency $\nu$.
Corresponding to the parameters used in Figure 6 we have $\eta=0.9$ and
$\nu_0=1.26$ GHz.
The best fit to Figure 6 yields a
power of $\eta=0.8$, which is quite close to the above prediction, considering
the fact
that the parameter $\Delta/(k_BT)=4$ used for the figure is far from the limit
of
dense levels.

\paragraph{}
For an electron turnstile using metal islands, the tunneling rates do not vary
with
time ($\delta=0$), and we should have $s\approx \Gamma_f^L/(2\nu)$ at $t=t_m$,
yielding the  same power law behavior of the trapping error in the asymptotic
range
$h_0\Gamma_f^L<<\nu<<\Gamma_f^L$.  However, the electron level density in a
metal
island is much larger than that in a quantum dot, and correspondingly the
tunneling rate
$\Gamma_f^L$ for a single level may be very small. It is therefore more
appropriate to consider
the regime of $\nu>>\Gamma_f^L$, where we may drop the second term on the right
hand side of Eq.(2).
Then a single simple equation for $P$ is obtained by summing over $\alpha$,
with the result
\bgeq
{dP\over dt}=(1-P) \Gamma_f^L\sum_\alpha a^{\alpha-\alpha_f} f_\alpha^L.
\label{ra-emt}
\edeq
Therefore, the trapping error relaxes exponentially with a single rate $\gamma$
given by
the quantity multiplying the factor $(1-P)$ on the right hand side of the above
equation.
The trapping error at $t=t_m$ goes exponentially as $e^{-\pi\gamma/\nu}$.  For
$a=1$ and at low
temperatures, this rate is the well known result $\gamma=(E_f^L-E_0)/(e^2
R_T)$,
where $E_f^L$ is the Fermi energy of the feeding lead (island), $E_0$ is the
energy of
the lowest available level in the trapping island, and $R_T=\Delta/(\Gamma_f^L
e^2)$
is the tunneling resistance.

\paragraph{}
However, the exponential relaxation of the trapping error does not mean that
the
distribution of probabilities on the levels also relax exponentially to
equilibrium.
In fact, with the known time dependence of $P$ substituted back into
Eq.(\ref{ra-emt}),
we found that
$P_\alpha$ has the long time limit of $\Gamma_f^L a^{\alpha-\alpha_f}
f_\alpha^L/\gamma$,
which is far from the Boltzmann distribution at equilibrium with the feeding
lead.
Indeed, the extra energy $\sum_\alpha P_\alpha (E_\alpha-E_0)$ of the trapped
particle
is found to be on the order of $E_f^L-E_0$ instead of $k_BT$ at equilibrium.
This failure of reaching equilibrium in the continuum limit can be understood.
The second term on the right hand side of Eq.(2) describes the process of
tunneling out
of the trapped particle, a necessary step for the redistribution of the
probabilities on the
levels, but we have dropped it due to the vanishing of $\Gamma_f^L$ in the
continuum limit.
In reality, such a term does exist although it is very small, and equilibrium
will be reached after a
very long time.  We will not look further into this, because this time scale
can be much longer than
an operation cycle of the device, or the time scales of inelastic processes
involving phonons or
background electrons.

\section{Hot Electron Relaxation Via Phonon Emission}
\paragraph{}
The above discussions have not taken into account interlevel transitions, which
may
facilitate the trapped electrons to attain equilibrium within the dot as well
as with the ambient.
In this section, we will estimate the transition rates due to phonon emission.
Both deformation
potential and piezoelectric coupling will be considered.  Particular attention
will be
given to the case of a quantum dot, in which the discreteness of the energy
levels plays
an important role \cite{Bock}.  For completeness, we have also considered the
continuum model,
which is appropriate for metal islands and for sufficiently large quantum dots.

\subsection{ Deformation Potential Coupling}
\paragraph{}
 Following common practice \cite{But}, we write the deformation potential
coupling as:
$H_d=\sum_{jk}E_{jk}\ep_{jk}$, where
$E_{jk}$ is the deformation potential tensor, and $\ep_{jk}$ is the symmetric
strain tensor
 ${1\over 2}[{{\partial u_j}\over{\partial x_k}}+
{{\partial u_k}\over{\partial x_j}}]$ [20].
The lattice displacement of the substrate
may be expanded in terms of the normal modes as
 ${\vec u}=\sum_{q,\lam}[{\hbar\over{2Mc_l|q|}}]^{1\over 2}
[\vec \xi_{q,\lam} e^{i\vec{q}\cdot\vec{R}}(a_{q,\lam}+a_{-q,\lam}^\dagger)]$,
where $M$ is the total
mass of the substrate, $c_l$ is the phonon speed, and
$a_{q\lambda}$ and $a_{q\lambda}^\dagger$ are the annihilation and
creation operators of a phonon with wave vector $\vec q$ and in mode
$\lambda$, respectively.  We have treated the lattice points $\vec R_j$ by a
continuous variable
$\vec R$, which is appropriate in the long wavelength limit.
The polarization vector $\vec{\xi_q}$ is required to
be real and satisfy $\vec{\xi_q}=-\vec{\xi_{-q}}$.

\paragraph{}
For an electron initially in state $|i>$ with energy $E_i$
and finally in state $|f>$ with energy $E_i+\delta E$,
the transition rate due to the deformation potential is obtained
using the Fermi's golden rule as:
\bgeq
\Gamma_{if}^{(d)}={\pi\over{Mc_l}}\sum_{q,\lam}
{{|\sum_{jk\lam}E_{jk} \xi_{q,\lam}^jq_k|^2}
\over q}|<i|e^{i\vec{q}\cdot\vec{R}}|f>|^2 \delta(\hbar c_lq-\delta E)
\bar{n}_B, \label{tr-de-qd,1}
\edeq
where  $\bar{n}_B=(n_B+{1\over 2}\pm{1\over 2})$
 with $n_B(\delta E)=[e^{{\delta E}\over{k_BT}}-1]^{-1}$, the positive sign is
for
$\delta E<0$ and negative for $\delta E>0$.
The stress tensor $E_{jk}$ may be reduced to a diagonal
form using its principal axes, and the sum involving $E_{jk}$ is
simplified to ${|\sum_{j\lam}E_{jj}(\xi_{q,\lam}^jq_j)|^2}$. In a spherically
symmetric conduction band valley, $E_{jj}=D$.  Then, Eq.(6) becomes
\bgeq
\Gamma_{if}^{(d)}={\pi\over{Mc_l}}D^2\sum_{q,\lam}
{{|\vec{\xi_{q,\lam}}\cdot\vec{q}|^2}
\over q}
|<i|e^{i\vec{q}\cdot\vec{R}}|f>|^2 \delta(\hbar c_l|q|-
\delta E)\bar{n}_B. \label{tr-de-qd,2}
\edeq
For a nonspherical band valley, the stress tensor components may be different
from each other, but we may still use Eq.(\ref{tr-de-qd,2})
by replacing the sum in
Eq.(\ref{tr-de-qd,1}) by its average over the direction of $\vec{q}$. We see
from
Eq.(\ref{tr-de-qd,2}) that only the longitudinal phonons, for which
$|{\vec{\xi_{q,\lam}}}\cdot\vec{q}|^2=q^2$, are involved in the transition.
Replacing the summation
$\sum_q$  by the integral ${V\over{8\pi^3}}\times \int d^3q$, where $V$ is the
volume  of the substrate, we derive that
\bgeq
\Gamma_{if}^{(d)}=({{\delta E}\over\hbar})^3{D^2\over{2\pi\hbar c_l^5\rho}}
{\overline M_{if}}\bar{n}_B,
\edeq
where $\rho$ is the mass density of the substrate,
$M_{if}=|<i|e^{i\vec{q}\cdot\vec{R}}|f>|^2$, and $\overline M_{if}$
is the average of transition matrix element $M_{if}$ over all possible
directions of
$\vec{q}$ for a given energy difference $\delta E$.
For GaAs, the deformation
potential $D=8.6$ eV, phonon speed $c_l=5136$ ms$^{-1}$, and mass density
$\rho=5.3079$ g/cm$^3$ \cite{Rode}, then Eq.(10) becomes
\bgeq
{\Gamma_{if}^{(d)}}=5.3{\times}10^{7}{\rm s^{-1}}\times ({\delta E\over 0.1{\rm
meV} })
^{3}{\bar M}\bar{n}_B.
\edeq

\subsection{Piezoelectric Coupling}
\paragraph{}
The piezoelectric potential $\phi_p$ satisfies the equation
\bgeq
\nabla^2_R\phi_p={1\over{\ep_0\ep_r}}\sum_{ijk}p_{ijk}
{{\partial\ep_{jk}}\over{\partial x_i}}.
\edeq
where $p_{ijk}$ are the piezoelectric constants.
 For detailed discussion of the piezoelectric
properties of GaAs crystal, the reader may consult Ref \cite{Nye}.
Since $\ep_{jk}$ can be expanded in terms of the normal modes as
$$\ep_{jk}={i\over 2}
\sum_{q,\lam}[{\hbar\over{2Mc_l|q|}}]^{1\over 2}[(\xi_{q,\lam}^jq_k+
\xi_{q,\lam}^kq_j)e^{i\vec{q}\cdot\vec{R}}a_{q\lam}-c.c.], \nonumber$$
the piezoelectric potential energy is obtained from Eq. (9)  as
\bgeq
e\phi_p=e\sum_{q,\lam}[{\hbar\over{2Mc_l|q|}}]^{1\over 2}
[\Xi_{q\lam}e^{i\vec{q}\cdot\vec{R}}a_{q\lam}+c.c.],
\edeq
where $\Xi_{q\lam}=\sum_{ijk}{{q_iq_k}\over q^2}\xi_{q,\lam}^jp_{ijk}$.

\paragraph{}
Applying the Fermi's golden rule again,
we obtain the transition rates due to the piezoelectric coupling as:
\bgeq
\Gamma_{if}^{(p)}={{\pi e^2c_l\bar{n}_B}\over{\ep_0\ep_rV}} \sum_q{1\over q}
<\ep_0\ep_r\sum_\lam{{|\Xi_{q\lam}|^2} \over{\rho c_l^2}}>
|<i|e^{i\vec{q}\cdot\vec{R}}|f>|^2 \delta(\hbar c_l|q|-
\delta E).
\edeq
Replacing $<\ep_0\ep_r\sum_\lam{{|\Xi_{q\lam}|^2} \over{\rho c_l^2}}>$ by the
piezoelectric constant ${\it p}^2$ \cite{But} and
the summation over $q$ by its corresponding integral, we obtain
\begin {eqnarray}
\Gamma_{if}^{(p)}&=&{e^2\over{2\pi\ep_0\ep_r}}{{\delta E}\over{\hbar^2c_l}}
{\it p}^2 {\overline M_{if}}\bar{n}_B \\
	&=&2.7\times 10^{10} {\rm s^{-1}} \times {\delta E\over 0.1 {\rm meV}}
{\overline M_{if}}\bar{n}_B.
\end {eqnarray}
In the second equality, we have used the values of
the piezoelectric constant $p=0.052$ and
the relative dielectric constant
$\varepsilon_{r}=12.91$ for GaAs \cite{Rode}.

\paragraph{}
The above result does not contain the effect of screening from the electrons
inside and
outside the quantum dot, and may be used as a first approximation when the
quantum dot
contains only a few electrons.  However, for a quantum dot of few thousand
{\AA}, hundreds
of electrons are present \cite{Kou:PRL,Kou:Z},
which considerably reduce the piezoelectric potential.
In such a case we may regard the electrons in the quantum dot as a 2DEG, for
which
the screening factor for the transition rate is given by $({a_{s}q_{\parallel}
})^2$, where
$a_{s}={2\pi\epsilon_0\epsilon_r \hbar^2\over e^2 m*}\sim 49.5{\rm \AA}$ is the
screening
length of a 2DEG in a GaAs heterostructure,
and $q_{\parallel}$ is the phonon wave number in the
plane of the 2DEG.  For a rough estimate, we may take $q_\parallel$ as
$n\pi/L$, where $n$ is
the number of levels between the initial and final states.
Taking a level spacing of $0.03$ meV, and dot size $L=4300$ \AA,
the screening factor may be written as $0.014 (\delta E/0.1{\ \rm meV})^2$
\cite{Kou:PRL,Kou:Z}.

\subsection {The Transition Matrix Element}

\paragraph{}
We now consider the transition matrix element $\bar{M}$.
Some detailed discussion of $M$ for different device geometry in
a 2DEG has been given in Ref. \cite{Bock}.
To understand the $q$ dependence of ${\bar M}$, we consider, for
instance, the case of a square dot  with
the transition $(k_x,k_y,k_z)=(10\pi/L_x, 10\pi/L_y, \pi/L_z)
)\rightarrow(10\pi/L_x, 9\pi/L_y,
\pi/L_z)$, where the $k$'s are the wave numbers of the electron state
$\sin(k_x x)\sin(k_y y)\sin(k_z z)$.
Using Eq.(4) of Bockelmann and Bastard \cite{Bock}, we obtain,
after some mathematical manipulations, the transition matrix element as
$M={{\sin^2 Q_{z}} \over Q_{z}^2}{{\sin^2 Q_{y}} \over Q_{y}^2}{{Q_{x}^2 \cos^2
Q_{x}} \over
 {[Q_{x}^2-({\pi \over 2})^2]^2}}$,
where we have introduced the reduced phonon wave numbers:
$Q_{x}=q_{x}{L_{x}/2}$, $Q_{y}=q_{y}{L_{y}/2}$, and $Q_{z}=q_{z}{L_{z}/2}$.

\paragraph{}
Now, it is instructive to give an estimate of the phonon wave length,
using the law of energy conservation: $\lambda=h c_l/\delta E$.
For $0.03$ meV $<\delta E<1$ meV, we find $\lambda\sim$230-7000 \AA.
Since the thickness of a 2DEG is only $\sim100$ \AA, we can set $M_{z}(q_z)=1$.
Then, the average of the transition matrix element over the directions of the
phonon
wave vector are found as:
${\bar M}{\simeq}{1 \over 3}({2 \over \pi})^{4}Q^2$ for $Q{\ll}1$
and $\simeq {\pi \over {4Q^2}}$ for $Q\gg1$,
where $Q=qL/2$, assuming $L_x=L_y=L$.
For transition energies $\delta E\ge 0.03$ meV (level spacing)
and a dot size of $L=4300$ \AA, we find $Q>2$.
Employing the formula for large $Q$, we
obtain ${\bar M}=2{\times}10^{-2}({\delta E\over 0.1{\rm meV} })^{-2}$.
This gives
\bgeq
{\Gamma_{if}^{(d)}}=10^6{\ \rm s^{-1}}\times {\delta E\over 0.1{\rm meV} }
\bar{n}_B,
\edeq
for deformation potential coupling, and
\bgeq
{\Gamma_{if}^{(p)}}=7.6\times 10^6{\ \rm s^{-1}} \times{ \delta E \over 0.1{\
\rm meV}}
\bar{n}_B \label{tr-pz-qd-1}
\edeq
for piezoelectric coupling (with screening).
It is seen that both deformation potential and piezoelectric couplings give a
transition rate of similar form, but the latter is more effective for GaAs.

Some comments are in order.  First, although we have considered a particular
pair of
initial and final states, the results also approximately hold for
other transitions in the dot.  Second, the large $Q$ form of the matrix element
can also be obtained  between a pair of plane wave states. Comparison with a
continuum model
will be discussed in subsection 3.5.  Third, when the wave length of the
emitted
phonons are longer than the size of the dot,  the matrix element in the small
$Q$
limit should be used, for which the transition rate $\Gamma_{if}\sim (\delta
E)^{5}$.
The scaling of $Q$ with the dot size is somewhat tricky.
For fixed $\delta E$, $Q$ decreases linearly with the size.
However, since the level spacing scales as $L^{-2}$, $Q$ increases as $L^{-1}$
with
decreasing the size, if the number of levels between the initial and final ones
is fixed.

\subsection {Effects on the Distributed Relaxation Rates}
\paragraph{}
To study the effect of phonon induced intradot transitions,
we add the rates determined by Eq.(\ref{tr-pz-qd-1}) to the right hand side of
Eq.(2):
\bgeq
{dP_{\alpha} \over dt}=  g_{\alpha}^L(t)
[(1-{\sum_{\alpha'} P_{\alpha'}})f_{\alpha}^L-P_{\alpha}(1-f_{\alpha}^L)]+
\sum_{j=1}^N [\Gamma_{j\alf}P_j-\Gamma_{\alf j}P_{\alf}],
\edeq
where $\Gamma_{j\alf}={\Gamma_{if}^{(p)}}$.
Since the phonon induced transitions leave the Boltzmann
distribution invariant, the equilibrium solution remains the same as before.
However, the non-equilibrium behavior is changed substantially.
To get an idea of how the
intradot transition rates modify the trapping process, we approximate
$g_f^L(t)$ by $\Gamma_f^L$ for $ |\omega t-t_{1\over 2}|< \sqrt{\pi\over
2\delta}$ and by $0$
otherwise.

\paragraph{}
The trapping error has the form of
Eq.(3) but with different eigenvalues $\{\lambda_j\}$ and eigenvectors.  As we
can
see from Table 1, dramatic changes occur in the slower modes represented
by the first four columns.
In particular, the lowest characteristic rate is now raised to
$\nu_1=10^{-2}\Gamma_f^L/\sqrt{2\pi\delta}\approx 12\times 10^6$ /s. This is
about the same as the phonon-induced rate for an electron
to relax from the level just below the Fermi energy of the left lead to the
lowest available level in the dot, which is $\Gamma_{if}$ at $\delta E=5\times
0.03$ meV.
We have also solved a similar case with ${\Delta\over {k_BT}}=2$, and find
that the lowest characteristic rate is raised to $\nu_1=8.4\times 10^6$ /s.
Thus, electron
phonon interaction helps to thermalize hot electrons in a Coulomb island.

\subsection{The Continuum Limit}
In this subsection we calculate the phonon induced
decay rate of a state in the Fermi liquid
model.  This is appropriate for a metal island in which the levels are densely
distributed.  The result for a quantum dot in the continuum limit is also
useful as a
reference for our understanding of the results calculated for the discrete
case.

Consider first a metal island, in which a state $\k$ at energy $\delta E$ above
the
Fermi sea is to decay into other states via phonon emission.  The rate at zero
temperature
is given by the Fermi golden rule as
\bgeq
\gamma_{eph}={2\pi\over \hbar}\sum_{k'}
|g_q|^2\delta(\hbar\omega_q-E_k+E_{k'}),
\edeq
where $\q=\k'-\k$ is the phonon wave vector and  $\omega_q=qc_l$ is the phonon
frequency.
The coupling function is taken as $|g_q|^2={\hbar\omega_q E_f\over 3nV}$, where
$n$ is the electron density and $V$ is the volume \cite{Ash}.  Replacing the
sum by a $k'$ integral
and carrying out the radial integration, we obtain
\bgeq
\gamma_{eph}={E_f\over 3n\hbar v_f (2\pi)^2}\int ds' \omega_q,
\edeq
where $ds'=2\pi k_f^2 \sin \theta\,d\theta$ is a Fermi surface element,
and $\delta E<<E_f$ is assumed.
Employing the relations $k_f^2\sin \theta\, d\theta=qdq$ and
$n=k_f^3/(3\pi^2)$, we carry out the
resulting
$q$ integral in the allowed range $0<q<\delta E/(\hbar c_l)$, yielding
\bgeq \gamma_{eph}={\pi (\delta E)^3\over 24\hbar E_f mc_l^2}, \edeq
where $m$ is the electron mass.  The cubic power dependence on energy can be
understood
by a simple power counting: 2 from the phase space of the phonons, and 1 from
the coupling
constant.

\paragraph{}

We now make some estimates.  For aluminum, $E_f=11.7$ eV and $c_l=6420$ m/s,
we have $\gamma_{eph}=0.072\times 10^6$ /s at $\delta E=0.1$ meV.  For copper,
$E_f=7$ eV and
$c_l=5010$ m/s, we have $\gamma_{eph}=0.2\times 10^6$ /s at $\delta E=0.1$ meV.

\paragraph{}

Next, consider a quantum dot in the continuum limit.  Using a similar method,
we obtain the decay rate
of a state due to emission of 3 dimensional phonons as:
\bgeq \gamma_{eph}=\int dq_z \int dl'{V|g_q|^2\over \hbar^2 v_f (2\pi)^2},
\edeq
where $q_z$ is the phonon wave number in the direction perpendicular to the
2DEG.
Replacing the Fermi line element $dl'$ by $dq_\parallel$, and carrying out the
angle integration
in the $(q_z,q_\parallel)$ plane, we have
\bgeq
\gamma_{eph}=\int \pi q dq{ V|g_q|^2\over \hbar^2 v_f (2\pi)^2}.
\edeq

\paragraph{}

For deformation potential coupling, we have $V|g_q|^2={\hbar D^2 q\over2\rho
c_l}$, yielding
\bgeq
\gamma_{eph}^d={D^2 (\delta E)^3\over 24\pi \rho c_l^4\hbar^4  v_f}.
\edeq
For a 2DEG in a GaAs/AlGaAs heterostructure, we may use $E_f=7$ meV and
 effective mass of an electron m=0.067m$_e$, yielding $v_f=1.9\times 10^5$ m/s.
 Using
the known parameters, we find $\gamma_{eph}=1.2\times 10^6$ /s at $\delta
E=0.1$ meV.

\paragraph{}

For piezoelectric coupling, we have $V|g_q|^2=\hbar e^2c_l p^2/(2
\epsilon_0\epsilon_r q)$.
This is to be screened  by a factor of $(q_\parallel a_s)^2$, which becomes
${1\over 2} (qa_s)^2$
after directional average in the $(q_z,q_\parallel)$ plane. The decay rate then
becomes
\bgeq
\gamma_{eph}={a_s^2 e^2 p^2 (\delta E)^3\over 48\pi v_f\hbar^4  c_l^2
\epsilon_0\epsilon_r},
\edeq
which yields $\gamma_{eph}=0.6\times 10^6$ /s at $\delta E=0.1$ meV.

\paragraph{}

Some comments are in order.  First, the cubic power dependence on energy
remains true
for the 2DEG, because we still used 3 dimensional phonons.  Second, in
subsection 3.3
the transition rate from a given initial state to a given final state was found
to be
$\sim \delta E$.  This is consistent with the $(\delta E)^3$ law of decay rate
found in the
present subsection, because here we have summed over the final states which
occupy a phase space
of order $(\delta E)^2$.  Third, for the example considered in the previous
subsections, there are
only  few available final states, so the transition rate found there also
corresponds roughly to the
decay rate of the initial state.  This rate is, however, considerablly larger
than the decay
rate predicted by the continuum model, illustrating the importance of
discreteness of the energy
levels.  Finally, formulas derived in this subsection may be
used for a crude estimate of the cooling rate of an electron gas at a
temperature $T_e$ higher than
the lattice temperature.  Take an electron at energy
$\delta E\sim k_BT_e$ above the Fermi energy, it will decay into lower states
with a rate $\sim
(k_BT_e)^3$ and with an energy transfer of order $k_BT_e$.  The number of such
electrons per unit
volume (area) is of order $k_BT_e$ times the density of states at the Fermi
energy.  The cooling rate
is therefore proportional to $T_e^5$ and to the volume (area) of the  Coulomb
island.  Quantitative
calculations of the cooling rates will be done in the next section.

\section{Self Heating in a Coulomb Island}
\paragraph{}
We now bring in an additional channel of relaxation of the tunneled-in
electron:
the inelastic scattering by other electrons in a Coulomb island.
Such scattering has two effects: (1) it limits the characteristic frequencies
of the device  to be above the typical rate $\gamma_{ee}$ of this channel, and
(2)
it thermalizes the tunneled-in electron with others, turning the extra energy
into
electronic heat.  If the thermalization process is fast enough, we may assign
an
electron temperature $T_e$ for the electron gas in the island, although it may
be
higher than the temperature of the lattice phonons in the substrate.
Phonons will be emitted from the electron gas to reduce the temperature
difference.
A steady state is reached when this cooling rate is balanced by the heating
rate
due to tunneling-in hot electrons (and holes).

\paragraph{}
In the following subsections, we first give a rough estimate
of the hot electron relaxation rates in both a metal island and a quantum dot
due to interactions with the background electrons there.  We then
calculate the cooling rates of thermalized electron gases  due to emission of
phonons; the results for metal islands of aluminum and copper and for a quantum
dot
will be compared to experimental measurements.
Finally, we show how the electron temperature (and the corresponding thermal
error) in a steady state
scales with the size and operation frequency for different devices.

\subsection{Relaxation Due to Background Electrons}
\paragraph{}

For a sufficiently clean and large metal island, we may use the Fermi liquid
formula $\gamma_{ee}={(\delta E)^2\over hE_f}$ to
estimate the inelastic relaxation rate of an electron of energy $\delta E$
above the Fermi energy $E_f$ of the metal island, where we have
assumed that the temperature of the background electrons is much smaller than
$\delta E$
\cite{Hod}.
The relaxation rate is found from this as $0.2\times 10^6$ /s for an Al island
with $\delta E=$0.1
meV.

\paragraph{}

The correction due to disorder enters in a somewhat complicated manner
depending on the
length scale $L_{\delta E}=\sqrt{\hbar v_f l\over \delta E}$, where $l$ is the
elastic
mean free path \cite{Alt}.  In order to have an idea of how big this length
scale can be, we take
$\delta E=0.1$ meV and $l=100$ {\AA}, then $L_{\delta E}\sim 4000$ {\AA} for
aluminum
or copper.
If $L_{\delta E}$ is smaller than the linear sizes of the island
in all directions, then an additive correction to the relaxation rate is given
approximately by $(hN_fL_{\delta E}^3)^{-1}\sim (\delta E)^{3/2}$, where $N_f$
is the density of
states at the Fermi energy.  This correction becomes larger than the Fermi
liquid result when $\delta
E$ is smaller  than the level spacing in a mean free volume $l^3$.  For $\delta
E=0.1$ meV, the
crossover occurs when $l$ is about 80 {\AA} for aluminum or copper.

\paragraph{}

On the other hand, if
$L_{\delta E}$ is larger than the linear sizes of the island  in all
directions, then $L_{\delta
E}^3$ in the above formula should be replaced by the  volume $V$ of the island.
 In this regime, the
correction to the relaxation rate is a constant in $\delta E$ and
becomes larger than the  Fermi liquid result at $\delta E=0.1$ meV when the
volume $V$
becomes smaller than $0.05\ \mu\rm m^3$.  Such an island volume is in fact
quite typical in
Coulomb blockackade experiments.  Finally,
the system can be in neither of these regimes if $L_{\delta
E}$ lis in between different length scales in different directions of the
island.  For example,
for a metal island of sheet geometry, the correction to the relaxation rate
goes as
$(hN_fL_{\delta E}^2 d)^{-1}\sim (\delta E)$, if $L_{\delta E}$  becomes larger
than the thickness
$d$ but still smaller than the other length scales of the island.  For more
details, the
reader is refered to \cite {Das}.

\paragraph{}
For a semiconductor quantum dot with a degenerate 2DEG, the Fermi liquid
formula for
the relaxation rate of an electron of energy $\delta E<<E_f$ reads:
\bgeq
{(\delta E)^2\over hE_f}[{1\over 4}+\ln 2+{1\over2}\ln { E_f\over \delta E}].
\edeq
We have obtained this result by following the method of Hodge {\it et al}
\cite{Hod},
and using the two dimensionally screened Coulomb interaction.
The factor including the logarithms is due to the two-dimensionality of the
problem,
and is numerically about 2 to 3 for $\delta E=1\sim 0.1$ meV,
where we have used $E_f=7$ meV for a GaAs/AlGaAs structure.  The relaxation
rate is then estimated
as $\gamma_{ee}=3.5\times 10^8$ /s for $\delta E=0.1$ meV, which is three
orders
magnitude higher than that for a metal island due to the much smaller
Fermi energy for the 2DEG.

\paragraph{}

Corrections due to disorder in the strictly 2D case is given by $\delta E\over
8 m v_f l$,
where $m$ is the effective mass \cite{Abr}.  (It is interesting to note that
this can also be written
in the suggestive form $(hN_fL_{\delta E}^2)^{-1}$ up to a factor of 4, where
$N_f={m\over \pi
\hbar^2}$ is the density of states in two dimensions.)   For $\delta E=0.1$ meV
and $E_f=7$ meV,
we find that the correction becomes larger than the Fermi liquid result when
$l<1700$ {\AA},
where we have taken $m$ to be 0.067 times the electron mass.

\subsection{Cooling Rate of a Metal Island}
\paragraph{}
We assume that the electron gas in a metal island has a well defined
temperature $T_e$,
which may be different from the lattice temperature $T_l$.  When a phonon of
momentum $\q$
and energy $\hbar\omega_q$ is emitted, an electron makes a transition from an
initial state $\k$ to a final state $\k'=\k-\q$, whose rate is given by the
Fermi golden rule:
$\Gamma_{{\vec q}, {\vec k},{{\vec k}^\prime}}={{2\pi}\over\hbar}|g_q|^2 \delta
(E_{\vec k}-\hbar\omega_q-E_{{\vec k}^\prime})[n_B+1]$, where $n_B$ is the
Bose distribution function for the phonons and $g_q$ is the coupling to be
specified later.
The rate of energy loss is
\bgeq
Q_e=2\sum_{{\vec k},{{\vec k}^\prime}}
\hbar\omega_q\Gamma_{{\vec q}, {\vec k},{{\vec k}^\prime}}
[f(E_k)(1-f(E_k^\prime))],
\edeq
where $f(E_k)$ is the Fermi distribution function for electrons, and
the factor of 2 comes from the spin degeneracy.  Following Ref. \cite{Zim}, we
replace the
summations by appropriate integrals over ${\vec k}$ and ${{\vec k}^\prime}$,
and carry out first the integration perpendicular to the Fermi surface,
yielding:
\bgeq
Q_e=
{V^2\over{16\pi^5\hbar^3v_f^2}}\int ds^\prime\int
ds{{(\hbar\omega_q)^2|g_q|^2}\over
{[e^{\hbar\omega_q/k_BT_e}-1][1-e^{-\hbar\omega_q/k_BT_l}]}},
\edeq
where $k_f$ and $v_f$ are the electron wave number and speed at the Fermi
surface,
 $ds$ and $ds^\prime$ are elements of the Fermi surface.

\paragraph{}
A similar consideration for the phonon absorption process yields an expression
for the
rate of energy gain, which is the same as the above except that
$T_e$ and $T_l$ have interchanged their places.  The net rate of energy loss
can then be written
as $Q=F(T_e)-F(T_l)$, where
\bgeq
F(T_e)={V^2\over{16\pi^5\hbar^3v_f^2}}\int ds^\prime\int
ds{{(\hbar\omega_q)^2|g_q|^2}\over
{e^{\hbar\omega_q/k_BT_e}-1}}.
\edeq
We can replace one of the surface integrals by the surface integral of the
relative variable
$\k-\k'$, then the other surface integral just yields a factor of the area of
the Fermi surface.
The relative surface integral can be further transformed to an integration over
$q$, by using
the relation $qdq=k_f^2sin\theta d\theta$,
where $\theta$ is the angle between $k$ and $k^\prime$ on the Fermi surface.
The result is
\bgeq
F(T_e)={{V^2k_f^2}\over{2\pi^3\hbar^3v_f^2}}\int_0^{2k_f}
qdq{{(\hbar\omega_q)^2|g_q|^2}\over
{e^{\hbar\omega_q/k_BT_e}-1}}.
\edeq
\paragraph{}
For simple estimates we may use $|g_q|^2={{\hbar\omega_q E_f}\over{3NZ}}$
\cite{Ash} for the
deformation potential coupling, where $NZ$ is the total number of electrons in
the
metal.  At low temperatures, we may further use $\omega_q=c_lq$,
and replace the upper limit of the integration by $\infty$.  The result is
the famous $T^5$ law for the cooling rate: $Q=\Sigma V (T_e^5-T_l^5)$, where
$\Sigma=24.9{m^2E_fk_B^5\over 6\pi^3\hbar^7nc_l^2}$,
with $n=NZ/V$ being the electron density.

\paragraph{}
We are now in a position to make some simple estimates.  For aluminum, we have
$E_f=11.7$ eV, $n=18.1\times 10^{28}/{\ \rm m^3}$, and $c_l=6420$ m/s, yielding
$\Sigma=0.1\ {\rm nW/K^5/\mu m^3}$.  This compares fairly well with the
measured value
of $\Sigma=0.2\ {\rm nW/K^5/\mu m^3}$ in Ref. \cite{Kau}, considering the
crudeness
of our model and the difficulty in the experiment.
For copper, we have $E_f=7$ eV,
$n=8.47\times 10^{28}/{\ \rm m^3}$, and $c_l=5010$ m/s, yielding
$\Sigma=0.2\ {\rm nW/K^5/\mu m^3}$.
Although this predicts correctly a larger cooling rate for copper than for
aluminum,
the numerical value is about an order of magnitude smaller than the
experimental result in Ref.
\cite{Nahum}.

\subsection{Cooling Rate of a Quantum Dot}
\paragraph{}
We now consider the cooling rate of a quantum dot made of a 2DEG in a
semiconductor heterostructure.  We assume that: (1) the dot is clean and small
enough
that impurity scattering may be neglected, and large enough that the
discreteness of
the energy levels may be ignored; (2) the electron gas form a degenerate Fermi
sea,
which is the case
when the electron temperature $T_e$ is below the liquid
helium temperature and the Fermi energy $E_f$ is at or above a few meV.
The cooling is achieved through coupling to phonons in the substrate, which
is held at a lower temperature $T_l$.

\paragraph{}
Following the method of the last section, the rate of energy loss by emitting
phonons is given by
\bgeq
Q_e=
{A^2\over{4\pi^3\hbar^3v_f^2}}\sum_{q_z}\int dl^\prime\int
dl{{(\hbar\omega_q)^2|g_q|^2}\over
{[e^{\hbar\omega_q/k_BT_e}-1][1-e^{-\hbar\omega_q/k_BT_l}]}},
\edeq
where as before $v_f$ is the Fermi velocity, and $g_q$ is the coupling
function.
The lower dimensionality of the electron gas is reflected as follows:
$A$ is the area of the dot, $dl$ and $dl^\prime$ are line elements of the Fermi
circle,
and we have lost a pair of factors of $1\over 2\pi$.  However, the phonons are
three dimensional, so $\q_\parallel=\k-\k'$, and we have the extra sum over
$q_z$.
Again, we may replace the line integrals by $2\pi k_f$ times $\int d
q_\parallel$.
Then, together with the $q_z$ sum, the latter integral may be written as
${L_z\over 2}\int qdq$, where $L_z$ is the thickness of the substrate in
the $z$ direction.  Therefore
\bgeq
Q_e=
{A^2 L_z k_f \over {4\pi^2\hbar^3v_f^2}}\int qdq
{{(\hbar\omega_q)^2|g_q|^2}\over
{[e^{\hbar\omega_q/k_BT_e}-1][1-e^{-\hbar\omega_q/k_BT_l}]}}.
\edeq
Combining with a similar expression for the rate of energy gain due to
phonon absorption, we can again write the net cooling rate as
$Q=F(T_e)-F(T_l)$, where
$\sigma A (T_e^5-T_l^5)$ at low temperatures, where the coefficient is given by
\bgeq
F(T)={A^2 L_z k_f \over 4\pi^2\hbar^3v_f^2} \int qdq { (\hbar\omega_q)^2|g_q|^2
\over
e^{\hbar\omega_q/k_BT}-1}.
\edeq

\paragraph{}
For deformation potential coupling, we have $AL_z|g_q|^2=\hbar q^2D^2/(2\rho
\omega_q)$,
yielding:
\bgeq
F(T)={AD^2k_f(k_BT)^5\over 8\pi^2\hbar^5\rho c_l^4 v_f^2}\int_0^\infty {x^4
dx\over e^x-1},
\edeq
at low temperatures.  This gives $Q=\sigma_d A(T_e^5-T_l^5)$, with
$\sigma_d=19\ {\rm fW/K^5/\mu m^2}$ for a 2DEG from GaAs heterostructure, where
we have used
the effective mass $m=0.067m_e$ and Fermi energy $E_f=7$ meV, together with
the material parameters for $\rho$, $D$, and $c_l$ (see Section 3).

\paragraph{}

For piezoelectric coupling, we have
$AL_z|g_q|^2=\hbar e^2c_l p^2/(2 \epsilon_0\epsilon_r q)$.  This is to be
screened
by a factor of $(q_\parallel a_s)^2$, which is ${1\over 2} (qa_s)^2$ after
directional
average. This yields a similar form for the net cooling rate as $Q=\sigma_p A
(T_e^5-T_l^5)$,
where
\bgeq
\sigma_p ={{3\zeta (5) k_fk_B^5}\over {2\pi^2\hbar^5 v_f^2 c_l^2}}
{a_s^2 e^2p^2\over\epsilon_0\epsilon_r},
\edeq
with $\zeta$ being the Riemman Zeta function, and $\zeta (5)\simeq1.03693$.
This is evaluated as $\sigma_p=10.2\ {\rm fW/K^5/\mu m^2}$, using the
parameters mentioned in the
last  paragraph.  The coefficient for the total cooling rate is
$\sigma=\sigma_d+\sigma_p=30\ {\rm
fW/K^5/\mu m^2}$.  This result agrees very well with the experimental result of
\cite{Mit} for a high mobility sample at $T_e>0.12$ K.  However, at lower
temperatures and for
low mobility samples, the theoretical result is too low to explain the
experiment.  More
theoretical work needs to be done to take the disorder effects into account.

\subsection{Scaling of Electron Temperature and Thermal Error}
\paragraph{}
In this subsection,
we will show how the electron temperature scales with the device operation
frequency
and size, and estimate its magnitude for different types of devices under
experimental conditions.  We will assume that $\nu<\gamma_{ee}$, where
$\gamma_{ee}$ is
the hot electron relaxation rate due to interaction with the background
electrons,
so that a meaningful electron temperature may be defined for the electron gas
in a quantum dot
or metal island, although it may be higher than the lattice temperature.
The electron temperature is determined by the condition
of power balance.  The heating power due to hot electron (hole) deposition
depends
on the relative magnitude of
$\nu$, $\gamma_{ee}$, and $\gamma_{eph}$, where $\gamma_{eph}$ is the
relaxation rate of a
tunneled-in electron via coupling to phonons. If $\gamma_{eph}<\nu
<\gamma_{ee}$,
phonon induced transitions can be neglected, and a hot electron relaxes by
coupling to the feeding lead for
a time $1/ \gamma_{ee}$ before being thermalized with the background electrons.
If $\nu<\gamma_{eph}<\gamma_{ee}$, the phonon channel is also involved in
the relaxation before thermalization.  If $\nu <\gamma_{ee}<\gamma_{eph}$, the
extra
energy of the tunneled-in electron is dissipated before heating
takes place.  For the sake of rough estimates, we will assume in the rest of
this subsection
that the energy relaxation of a hot electron
due to tunneling and phonon emission may be neglected.
The cooling power due to net phonon emission from the thermalized and hot
electron gas
has been studied in the previous subsections.

\paragraph{}
Consider an electron turnstile made of  metal islands with operating
frequency $\nu$, volume $V$,
 and Coulomb charging energy $U$. The heating power due to the
deposition of hot electrons and holes in the island is $\simeq \nu U$.
Assuming that the lattice temperature $T_l$ satisfies $T_l^5\ll T_e^5$, we
obtain the cooling power as $\simeq \Sigma V
T_e^5$,  where $\Sigma$ is a dissipation constant $\simeq 0.2$ nWK$^{-5}{\mu
m}^{-3}$ \cite{Kau}.  In the steady state, we have $T_e=({{\nu U}\over{\Sigma
V}})^{1\over 5}$.
Using the facts that $U={e^2\over
{2C}}$ and $C\sim V^{1\over 3}$, we obtain that $T_e\sim\nu^{1\over 5}U^{4\over
5}$.
Hence, the thermal exponent that determines the thermal error is found as
\bgeq
{U\over{k_BT_e}}=\eta({U\over{h\nu}})^{1\over 5},
\edeq
where $\eta=({{2h\Sigma V U^3}\over{k_B^5}})^{1\over 5}$
is a dimensionless number, independent
of the operating frequency and the volume of the island because $U$ is
inversely proportional
to the linear size of the island.
Using the known experimental values $V=0.1\ {\mu \rm m}^3$ and
$C=1$ fF and $\nu=10$ MHz, we find $U=0.07$ meV, $\eta=2$, and $T_e=80$ mK.
The electron
temperature is  much higher than the ambient temperature $T=10$ mK.

\paragraph{}
A similar result can be obtained for a quantum dot device, but the parameter
$\eta$ is
given by $\eta=({{2h\sigma A U^3}\over{k_B^5}})^{1\over 5}$.
Note that $\eta$ now weakly depends
on the linear size of the dot but not on the operating frequency.
For an estimate,
we use the data in Ref. \cite{Kou:PRL}, where $A=0.5\ \mu\rm m^2$, $C=0.24$ fF,
yielding $T_e=340$ mK
at $\nu=10$ MHz, which is much higher than the reported $10$ mK ambient
temperature.
$\eta$ is found as $\sim 1.9$ for the given data.
\paragraph{}

In an electron pump, the amount of heating in a cycle is not simply given by
the charging energy
because  a tunneling stage is controlled by ramping a gate voltage. For a
3-junction pump (See Figure
3),  the heating energy can be derived as
\bgeq
\Delta E={U\over 2}(64R_TC\nu)^{1\over 2},
\edeq
according to Eqs.(35) and (38) of Ref. \cite{Jensen}. Again, we write the
thermal exponent
as $\eta({U\over{h\nu}})^{1\over 5}$ and find that
\bgeq
\eta=({{2h\Sigma V U^3}\over{k_B^5}})^{1\over 5}(64R_TC\nu)^{1\over{10}} ,
\edeq
which weakly depends on $\nu$ and $R_T$.  For typical values of $R_T=100\ \rm
k\Omega$,
$C=1$ fF, and $\nu=10$ MHz, we have $\eta\simeq 1.3$.

\paragraph{}
In view of the above scaling results,
we write the thermal error of these devices $\exp (-{U\over{k_BT_e}})$ in the
more
convenient form of
$10^{-\eta^*(U/\nu)^{1/5}}$, where $U$ is measured in units of meV, and $\nu$
in 10MHz.
The conversion relation is $\eta^*=3.26\eta$. The thermal errors are then on
the
order of $10^{-5.1}$, $10^{-6.2}$, and $10^{-6.75}$ for the above typical
examples
respectively.
These estimates seem to compare very well with the errors found in experiments,
but the reader
is warned not to take the numerical values too seriously because of the crude
nature
of our estimates.  The scaling forms should be more generally valid.

\section{Conclusion}
\paragraph{}

In summary, we have presented a detailed study and overview of the
nonequilibrium effects
and self heating in single electron transfer devices which are primarily
based on the Coulomb blockade
effect.  The Coulomb charging energy is taken to be much larger than the
temperature (electronic or
lattice), but the level spacing in the excitation spectrum is not necessarily
so.
We have considered three types of devices: quantum dot electron turnstile,
metal island electron
turnstile, and metal island electron pump.  In the turnstiles, a hot electron
may be
deposited in a quantum dot or metal island with an extra energy on the order of
the
Coulomb charging energy.  In the electron pump, the extra energy of the hot
electron depends on
the rate of pumping.

\paragraph{}

We have considered three major channels of hot electron relaxation: (1) via
tunneling, (2)
emitting phonons, and (3) exciting background electrons.  Three distinct
regimes are identified:
non-equilibrium ($\gamma_{ee},\gamma_{eph}<\nu$), self heating or partial
equilibrium
($\gamma_{eph},\nu<\gamma_{ee}$), and equilibrium
($\gamma_{ee},\nu<\gamma_{eph}$), see Figure 4
for an illustration.  For the quantum dot device used in \cite{Kou:PRL,Kou:Z},
$\gamma_{eph}$
is estimated to be about $10^7$ /s, while $\gamma_{ee}$ is on the order of
$10^{9}$ /s, so that
it lies in the self heating regime for practical operating freqencies.
However, $\gamma_{ee}$ may be reduced by depleting the background electrons in
the dot, rendering
the system to be in the other regimes.
For a metal island, both $\gamma_{eph}$ and
$\gamma_{ee}$ are found to be on the order of $10^5-10^6$ /s, so for $\nu>10^6$
Hz the system is
in the nonequilibrium regime, while for slower frequency the system lies near
the boundary
between the equilibrium and self heating regimes.

\paragraph{}

In the nonequilibrium regime, a hot electron relaxes only through tunneling
back and forth to the
feeding lead (or island).  A set of characteristic rates are found for the hot
electron relaxation.
Those rates corresponding to  the trapping levels below the Fermi energy of the
feeding lead are
exponentially small at low temperatures.  This is understood as due to the
Pauli exclusion in the
lead, which is more  and more effective at lower temperatures in
blocking the tunneling-out process.  A distributed set of characteristic rates
give rise to
an overall power law frequency dependence in the trapping error, which is
closely related to
the electron transfer error.  For energy independent couplings between the
levels and the lead, the
power law is linear up to  a logarithmic factor.  If the higher levels couple
more strongly to
the lead, a sublinear power law is obtained.  In the continuum limit, the
tunneling-out process
is nonoperative, and therefore a trapped hot electron is unable to relax.

\paragraph{}

We then considered the relaxation due to phonon emission for a hot
electron in a quantum dot, and studied its effect on the characteristic rates
due to tunneling.
Perhaps nonsurprisingly, those slow characteristic rates are found to be pushed
up to and above the
relaxation rate $\gamma_{eph}$ due to phonon emission.  When the device
operation frequency $\nu$ is slower than
$\gamma_{eph}$, the hot electron has enough time to reach equilibrium with the
phonons (the
ambient), with a residual trapping error exponentially small as a
function of $\nu$.  For completeness, we have also considered a continuum model
for metal island
and large quantum dots.

\paragraph{}

There is then the self heating regime, in which the hot electron quickly
reaches equilibrium
with the background electrons, but its extra energy is turned into electronic
heat, elevating
the electron temperature $T_e$ above the lattice temperature $T_l$.
We have rederived the formula $Q=\Sigma V (T_e^5-T_l^5)$ for the energy
dissipation rate (cooling power) to the phonons from a hot electron gas in  a
metal island of volume
$V$,  and obtained a simple expression for the coefficient $\Sigma$.  The
coefficient is
estimated  to be 0.1 nW/K$^5/\mu$m$^3$ for aluminum and 0.2 nW/K$^5/\mu$m$^3$
for copper.  These
results are in reasonable agreement with experiments \cite{Kau}.  A similar
expression, $Q=\sigma A
(T_e^5-T_l^5)$, has also been obtained for a quantum dot of area $A$, where the
coefficient
is estimated
as  0.03 pW/K$^5/\mu$m$^2$ for coupling to three dimensional phonons,
with the result in fairly good agreement with the experimental finding of
\cite{Mit} for a high
mobility sample.

\paragraph{}

The elevated electron temperature is obtained by balancing the cooling power
with
the heating power, and is found to be $\sim 100$ mK for a metal island and
$\sim 300$ mK for a quantum dot under typical experimental conditions, while
the ambient
temperature is much lower.
Different scaling behaviors have been obtained for the thermal
exponent $U/(k_BT_e)$ as function of operation frequency and size for different
type
of devices.  The estimated thermal errors are consistent with experimental
results.

\paragraph{}
There are two ways to solve the nonequilibrium and self heating problem.
One way is to increase the
rate of heat dissipation from the dot or metal island.  This can be done to
some extent by
device scaling and by
engineering the shape of the device structure in such a way that the dot area
or island
volume is maximized for a given capacitance. The other way is to reduce heat
production.
The electron pump is better than the
turnstile precisely because of less heat production, especially so at low
operation
frequencies \cite{Urb}.  One may also consider making a quantum charge pump (or
electron load-lock), which utilizes discrete electron levels to trap electrons,
and has the advantage of producing essentially no heat in the dot \cite{Niu}.

\paragraph{}

Finally, we emphasize that although our study were focused on single electron
transfer
devices, our results may also apply to other types of Coulomb blockade devices,
such as
electrometers \cite{Kau} and single electron transistors \cite{Kas}.  The role
of
device frequency $\nu$ is now played by the electron current divided by the
electron
charge.

\section*{Acknowledgment}
The authors wish to thank J. M. Martinis, L. P. Kouwenhoven, H. L. Edwards, A.
L. de Lozanne and
W. P. Kirk  for many valuable discussions.
This work was supported by a Precision Measurement Grant from NIST and the
Welch Foundation.

\clearpage
\section*{Appendix: Non-Markovian Effects}

\paragraph{}

The trapping errors displayed in Figure 6 are very small.  One wonders if the
semiclassical rate equations (1) are accurate enough for the prediction of
these
small quantities.  One possible correction to the semiclassical rate equations
comes from the non-Markovian effect.
This question was investigated in detail in Ref. \cite{Liu.dis}, and here we
give
a short summary:

\paragraph{}

 Starting from the non-Markovian master equation (7) in  Ref. \cite{Bruder},
a set of differential-integral equations for the level
occupation probabilities were derived.
These equations were then approximated by the semiclassical rate
equations plus leading order corrections.
The trapping errors were then calculated with the improved rate equations,
with the results (under conditions of figure 6):
$6.27\times 10^{-2}$,
$7.71\times 10^{-3}, 3.27\times 10^{-3}, 4.35\times 10^{-4}, 1.73\times
10^{-4},
9.79\times 10^{-6}$ for operating frequencies $\nu=$ 1 GHz, 100 MHz, 50 MHz,
10 MHz,  1.25 MHz, and 0.125 MHz respectively.
These are in good agreement with  the corresponding semiclassical results
$5.88\times 10^{-2}$,
$7.12\times 10^{-3}, 2.97\times 10^{-3}, 3.85\times 10^{-4}, 1.69\times
10^{-4}$, and $9.37\times
10^{-6}$.

\clearpage
\section*{Figure Captions}
\paragraph{}
Fig. 1.  Electron trapping in a quantum dot electron turnstile, which consists
of
a quantum dot (middle), two leads (with Fermi levels $E_f^L$ an $E_f^R$), and
energy
barriers separating them.  The Coulomb charging energy in the dot is denoted as
$U$,
and the level spacing as $\Delta E$.
With the left barrier lowered and under the biases shown in the figure,
it is energetically favorable for one and only one electron to tunnel
into the dot from the left lead, but the final state can be any one
of the several available states in the dot.

\paragraph{}
Fig. 2. The circuit for a metal island electron turnstile device.  The metal
islands
lie in between the tunneling junctions of capacitances $C$.  $C_g$ and $V_g$
are the
gate capacitance and votage, respectively.

\paragraph{}
Fig. 3.  The circuit for a metal island electron pump.
Each island is now separately gated.
Electron tunneling through each junction is controled by ramping the bias
accross it.

\paragraph{}

Fig. 4.  Three different regimes in the  parameter space of $\gamma_{eph}$ and
$\gamma_{ee}$,
the relaxation rates of a hot electron through emitting phonons and exciting
the background
electrons.  In the equilibrium regime, a trapped hot electron may achieve
equilibrium
with the background electrons and the lattice phonons before the trapping cycle
is finished.
In the self heating regime, the hot electrons may quickly thermalize with the
background electrons,
but with the latter heated to a temperature above the ambient lattice
temperature.
In the non-equilibrium regime, a trapped hot electron can relax only through
tunneling
back and forth to the feeding leed with a set of distributed characteristic
rates, the lower part of which can be extremely slow due to Pauli exclusion.

\paragraph{}

Fig. 5. Trapping probability $P(t)$  vs. $\omega t$ (solid line).  Also shown
are the
currents  $I_L$ (dot-dashed line) and $I_R$ (dashed line) in units of $\omega$.
The arrows indicate the times at which $P(t)$ achieves its minimum, half value,
and
maximum.

\paragraph{}

Fig. 6. The solid curve is the approximate theoretical prediction from Eq. (3),
which fits very well the results of an accurate numerical similation
(diamonds).
The exponentials indicate the behaviors of the downward curving
segments, with $\nu_2=14\times 10^3$ /s,  $\nu_3=1.2\times 10^6$ /s, and
$\nu_4=96\times 10^6$ /s.
The overall behavior is power law like, with $1-Max(P)\sim \nu^{0.8}$.

\clearpage
\section*{Table Caption}
Table 1. Two sets of egienvalues $\lambda_j$ in the absence and presence of
phonon
induced transitions.  The parameters taken are $N=10$, $\delta=10$,
${\Delta\over{k_BT}}=4$, and  $\Gamma_f^L=\Gamma_f^R=10^{10}$ /s.

\end{document}